# Categorizing Flight Paths using Data Visualization and Clustering Methodologies

Yifan Song, Keyang Yu
Department of Computer Science & Engineering
The Ohio State University
Columbus, OH, USA
song.1221@osu.edu, yu.2049@osu.edu

Seth Young, Ph.D.
Dept. of Civil, Environmental, and Geodetic
Engineering / Center for Aviation Studies
The Ohio State University
Columbus, OH, USA
young.1460@osu.edu

*Abstract—* **This work leverages the U.S. Federal Aviation Administration's Traffic Flow Management System dataset and *DV8*, a recently developed tool for highly interactive visualization of air traffic data, to develop clustering algorithms for categorizing air traffic by their varying flight paths. Two clustering methodologies, a spatial-based geographic distance model, and a vector-based cosine similarity model, are demonstrated and compared for their clustering effectiveness. Examples of their applications reveal successful, realistic clustering based on automated clustering result determination and human-in-the-loop processes, with geographic distance algorithms performing better for enroute portions of flight paths and cosine similarity algorithms performing better for near-terminal operations, such as arrival paths. A point extraction technique is applied to improve computation efficiency.**

## I. INTRODUCTION

It is highly common for flights within the United States' national airspace system to fly flight paths between their origins and destinations that vary, or deviate, from their planned flight routes. While planned flight routes are quite predictable, most often following jet routes and departure and arrival procedures, all of which are published as standard by the Federal Aviation Administration, the actual paths flown are more stochastic, with less understanding of any patterns among paths flown between any given origin and destination airport.

The high-level-variability of flight paths over planned routes leads to inefficiencies in travel time, fuel burn, and aircraft utilization. If there was a method to measure the variability of flight paths as a function of their planned routes, the industry may be able to focus their efforts on creating efficiencies for these routes. One necessary element of such an analysis would be to have some robust method of categorizing a spectrum of independent flight paths into clusters. This need serves as the motivation for this research.

This paper describes a flight path clustering methodology that leverages the Federal Aviation Administration's NextGen System Wide Information Management (SWIM) Traffic Flow Management System (TFMS) data feed, a software tool known as *DV8* used for fast-time visualization of TFMS data, and two clustering distance models for determining route clusters based on these visualizations, leveraging both automated and human-in-the-loop (HITL) determined clustering procedures.

### A. Background / Literature Review

Flight path (air trajectory) analysis is a crucial contributor to air traffic safety and efficiency. With growth in air traffic data feeds and the development of machine learning (ML) techniques, more and more ML methodologies have been introduced in flight route (trajectory) analysis. One of the early approaches is from de Leege et. al. to predict the landing trajectory in the landing process by training a generalized linear model [1]. Like de Leege, most traditional works used historical data in the prediction process. Recently, with the development in deep learning techniques, research as exampled by Liu and Hansen [2] consider a new model using, for example, convolutional and recurrent neural networks to predict the flight trajectories in a 4D, to include location, time, and the weather conditions in the vicinity.

The other approach to flight path analysis with machine learning is cluster analysis. There is a rich set of literature in this area. Gariel, et. al., were among the first to apply trajectory clustering into the aviation environment [3]. More recently, the use of newly available data and DBSCAN algorithm to cluster air traffic flows both in the enroute airspace (Duong, et. al. [4]), and in the airport terminal environment (Olive, et. al [5]). Another popular clustering algorithm, known as agglomerative hierarchical clustering, is used by Pusadan et. al. [6] to detect outlier flights with abnormal waypoints and Bombelli et. al. to identify and approximate well-traveled route in air strategic planning [7].

As an alternative to previous approaches, we propose a flight path clustering framework based on a flight data analysis and visualization tool known as *DV8*, developed by Young, et. al. [8]. Our clustering function focuses on the efficiencies associated with interactive analysis, allowing a user to query, visualize, and ultimately cluster hundreds of flights in a matter of seconds. Our framework also provides the flexibility for users to select different cluster models and thresholds to generate different clustering results based on different analysis needs and display the visualization and statistic results to let the users validate themselves.

## II. DATA AND VISUALIZATION MODEL

### A. FAA SWIM TFMS Data

The data used for this project emanates from the Federal Aviation Administration's SWIM program. Through the SWIM program the FAA disseminates large data sets describing the current air traffic environment in the U.S. National Airspace System [9]. The SWIM program offers several data sets. This project leverages the TFMS data set. TFMS provides real-time messaging of flight status (Such as flight plan created or modified, as well as other information including origin and destination airports, aircraft type, and aircraft operator), positioning, (latitude, longitude, speed, and altitude of aircraft in flight) and planned flight routes. The TFMS dataset disseminates its messaging through a .xml stream that may be ingested by external users upon completing the required FAA connection protocols [10].

### B. DV8 Visualization Tool

To date, existing analysis tools for visualizing and analyzing large aviation data such as SWIM TFMS are in many ways inefficient: they are either time consuming, taking minutes even hours to query and process data in a static environment or visualizing in a one-shot manner. As authored by, Omidvar-Tehrani, et. al. [11], DV8, an interactive aviation data visualization and analysis framework, was developed to accommodate a series of query-response (interactive) analysis with historical data and dynamic comparison time-wise and location-wise. With novel techniques in caching, sampling (quantized granularity) and indexing, DV8 can visualize and interpret the results in sub-seconds. DV8 is implemented in Python as the computation engine with a PostgreSQL database and a web interface front-end in JavaScript. Being a web service, DV8 has no prerequisites and can be executed with a browser on any platform. Fig. 1 illustrates the DV8 system architecture.

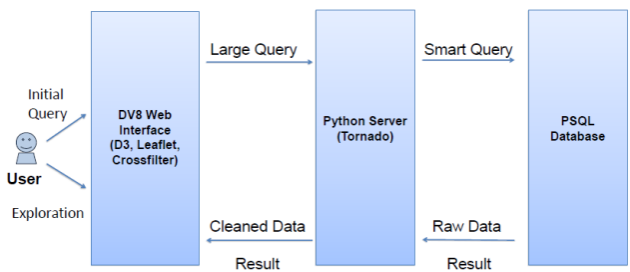

*Figure 1: DV8 System Architecture*

Fig. 2 illustrates an example of the DV8 visualization of all flights from Columbus (CMH) to New York (JFK) in May 2014. DV8 provides multiple visualization tools including color coding, dynamic filtering and toggle drawing. DV8 also provides descriptive statistics of queried flights, as well as the ability for users to focus on a single flight and download processed data. From all these features, users can easily recognize flight patterns interactively via visualization and do a secondary analysis or development based on different needs. One of the most noticeable flight patterns is the flight path. From the figure, there are certainly multiple flight path choices and some outlier routes. This visualization motivates the need to develop clustering algorithms to categorize flights by groups of routes. Thus, this is the motivation for the research.

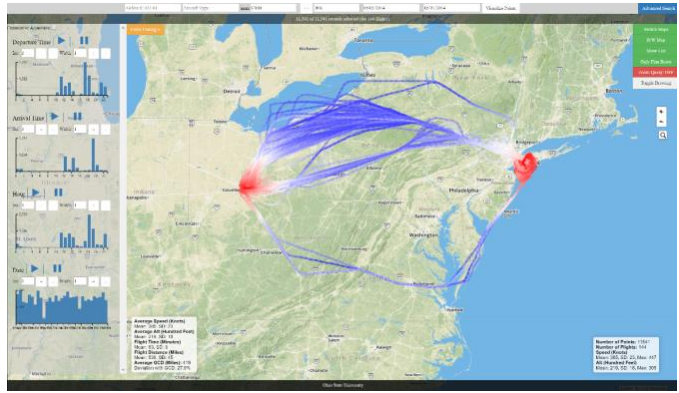

*Figure 2: DV8 Visualization, CMH-JFK, May 2014*

## III. ROUTE CLUSTERING METHODOLOGIES

As discussed in Section II, DV8 provides the ability to visualize multiple flight paths flown between two given airports. This visualization allows for easy visually and intuitively categorizing and clustering flight paths. However, for a human to do so for large numbers of flights over many airport-to-airport pairs, would be highly cumbersome. Using a machine learning algorithm to perform the categorization of flight paths would automate this process and provide more information. One of the most direct and useful machine learning applications for flight path learning is cluster analysis.

### A. Hierarchical Clustering

Cluster analysis is an unsupervised learning task that is used to detect potential groupings within a dataset without the need to provide corresponding categorizations a-priori [12]. Cluster analysis would assign the whole dataset into a number of groups that all data points within the same group are closer (more similar) to its counterparts in the same group and more distant (dissimilar) to the data points in the other groups. By using a clustering algorithm, our application categorizes flights into several groups according to their flight paths and detect possible outliers.

Our cluster analysis function is implemented by a hierarchical clustering algorithm, known as an agglomerative hierarchical clustering [13]. Fig. 3 provides an illustration of the algorithm.

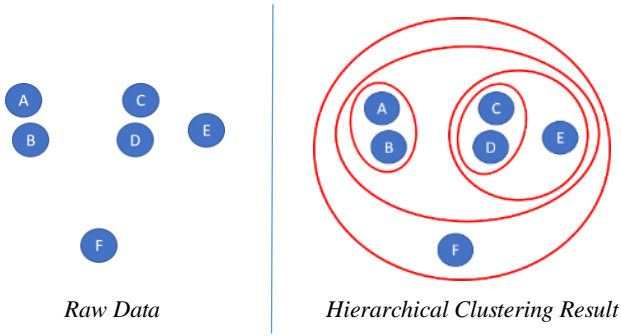

*Raw Data* | *Hierarchical Clustering Result*

*Figure 3: Hierarchical Clustering*

In agglomerative hierarchical clustering algorithm, initially, all data points are considered as an individual cluster. Then based on a specific distance metric, the algorithm will compute the distance matrix for all pairs of data points. It will merge the two nearest clusters and update the distance matrix. In each iteration, two closest clusters will be merged until there is only a single cluster remains or certain threshold is met. The pseudo code of the algorithm is as follows:

*Let each data point be a cluster*

*Compute the distance matrix between each cluster*

**Repeat**

    *Merge two closest clusters as one*

    *Update the distance matrix*

**Until** *certain threshold or only one cluster remaining*

The hierarchical clustering algorithm can often be visualized as a dendrogram, a tree-like diagram that records the sequences of merges or splits, as illustrated in Fig. 4.

The advantages of using hierarchical clustering are that: 1) hierarchical clustering does not depend a priori on a particular number of clusters; 2) the result, a specific number of clusters or a particular threshold, can be obtained by "cutting" the dendrogram at the proper level which provides more flexibility for analysis.

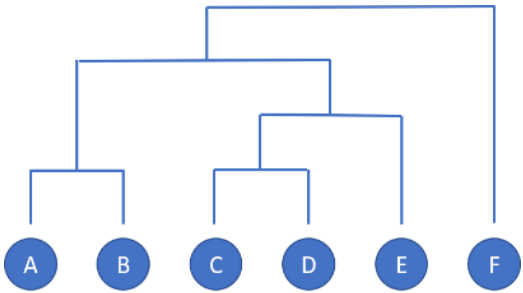

*Figure 4: Dendrogram*

To obtain the best "cutting level", Rousseeuw [14] notes that the use of a Silhouette score provides a metric for evaluating the level of cutting that best represents optimal clustering. To determine the best clustering result, our algorithm will loop through the dendrogram and find the level of cutting that yields the best Silhouette score. A Silhouette value of one sample is a measurement of how well that sample is clustered. More specifically, it is defined as:

$$(b - a) \, / \, max(a, b) \qquad (1)$$

Where $a$ is the mean intra-cluster distance and $b$ is the mean nearest-cluster distance. Thus, a higher Silhouette value for a sample indicates that the sample is better matched to its own cluster and worse matched to neighboring clusters. The Silhouette score is then the mean of Silhouette values over all samples.

### B. Applied Distance Clustering Models

Our application considers two distance models for comparative clustering computations: a spatial-based metric based on geographic distance, and a vector-based metric using the concept of cosine similarity.

#### 1) Geographic Distance: Spatial-based Model

The spatial based algorithm is based on geographic distance, more specifically, great circle distance (GCD). GCD is used to measure the distance between two points on the surface of Earth from latitude and longitude information. The Geographic Distance can be understood as an alternative of Euclidean Distance which is one of the most common distance functions, also often applied in air trajectory clustering, as previously referenced in [6]. The agglomerative hierarchical clustering algorithm takes the average of great circle distances between each pair of points for a given pair of flights. The formula is as follows:

$$d = \frac{1}{N} \sum_{i}^{N} GCD_i$$

$$where \;\; GCD_i = R \cdot arccos \begin{pmatrix} sin(a_{1i}) \, sin(a_{2i}) \, + \\ cos(a_{1i}) \, cos(a_{2i}) \, cos(|b_{1i} - b_{2i}|) \end{pmatrix} \quad (2)$$

Where $R$ is the earth's radius, $a_{ij}$ and $b_{ij}$ represents each latitude and longitude points from flight $i$ point $j$ respectively, the distance between two flights $d$ is then calculated by taking the average of GCD between each pair of points. The geographic distance can be any real-value, in our case nautical miles (nm).

#### 2) Cosine Similarity: Vector-based Model

Our second metric developed is based on a cosine similarity vector-based algorithm [15]. While the intuition of Geographic Distance is to measure the physical distance between two flights while the vector-based model is to measure the difference in their flying directions. For every two consecutive data points in a flight route, we subtract their longitude and latitude to get a vector of flight direction. We then calculate cosine similarity based on every pair of vectors to measure the direction similarity between two flight routes. The agglomerative hierarchical clustering algorithm takes the average of cosine similarity value between each pair of points extracted. The distance is calculated by one subtracting the average cosine similarity. Equations (3) through (6) describe how the distance is calculated.

$$X_i = (a_{i+1} - a_i, b_{i+1} - b_i) \qquad (3)$$

$$s_i = \frac{A_i \cdot B_i}{\|A_i\| \|B_i\|} \qquad (4)$$

$$S = \frac{1}{N} \sum_i^N s_i \qquad (5)$$

$$d = 1 - S \qquad (6)$$

Equation (3) shows how the vector is generated. Equations (4) and (5) are used to calculate cosine similarity for one pair of vectors where $A_i$ is the $i_{th}$ vector in the first flight and $B_i$ is the $i_{th}$ vector in the second flight. The distance between two flights is calculated by 1 – the average cosine similarity among pairwise vectors (6). The cosine similarity is in the range of [-1, 1] while 1 representing two vectors with exact same direction, -1 for exact opposite direction, and 0 for perpendicular vectors.

## IV. EXAMPLE APPLICATIONS OF CLUSTERING MODELS

In this section, we will show some actual examples of how our clustering models perform. We will show how different our two proposed methods perform in different circumstances with Silhouette score and visualization results.

### A. Visualization & Statistical Clustering Result

The first example comes from the following query: CMH (Columbus) to ATL (Atlanta) in the period of 06/01/2014 – 06/22/2014. In this example, two potential clusters are visualized. Most flights are in the east cluster that fly relatively similarly from CMH to ATL; the other small proportion of flights belonging to the west cluster choose to fly with greater variability. We will demonstrate how our agglomerative hierarchical clustering algorithm would perform to categorize such flight queries.

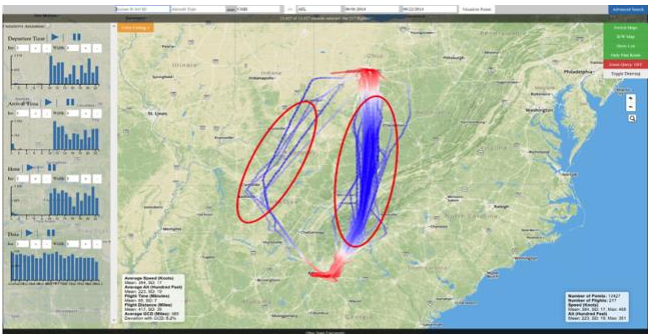

*Figure 5: Flights from CMH to ATL, 6/1/2014 - 6/22/2014*

Fig. 6 demonstrates the result of our clustering function using each algorithm, with a geographic distance (on the left) and a cosine similarity (on the right) as the algorithm. The result seems to have a perfect match with our intuitive manual grouping.

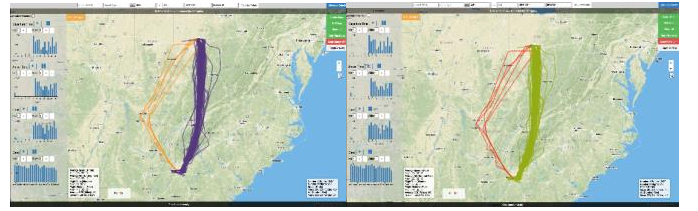

*Figure 6: Clustering of CMH-ATL flights (Left: Geographical Distance Silhouette Score 0.84 / Right: Cosine Similarity Silhouette Score 0.85)*

Table 1 describes relevant performance statistics of the two clusters generated by the agglomerative hierarchical clustering algorithm.

Table 1: Descriptive statistics of CMH-ATL categorized routes

| Cluster 1 | | Cluster 2 | |
|---|---|---|---|
| Number of Points: 13036 | Number of Flights: 212 | Number of Points: 391 | Number of Flights: 5 |
| **Speed (Knots)** | | **Speed (Knots)** | |
| Mean: 382 | SD: 17 | Mean: 370 | SD: 15 |
| **Altitude (Hundred Feet)** | | **Altitude (Hundred Feet)** | |
| Mean: 223 | SD: 19 | Mean: 232 | SD: 12 |
| **Flight Distance (nm)** | | **Flight Distance (nm)** | |
| Mean: 420 | SD: 45 | Mean: 415 | SD: 35 |
| **Deviation with GCD** | | **Deviation with GCD** | |
| 10.0% | | 7.8% | |

The above statistics demonstrate the effect of using our clustering function for generating the desired clustering results. Our clustering algorithm is able to identify a group of flights with a larger deviation from GCD (great circle distance) between two airports. The clustering result also demonstrates one cluster with a larger variation in flight distance and lower average speed.

### B. Comparing Each Model

While each model has its strengths, each encounter applications in which they are weak. In some cases, both models reveal feasible results for the same group of flights. For example, referring back to Fig. 6, a reasonable clustering of flights between CMH and ATL was achieved by both geographical distance and cosine similarity methods. But in other cases, it is found that one model may provide feasible results, while the other may not, as demonstrated in the previous two examples.

#### 1) Example where Geographic Distance Model is Preferred

In this next example, we demonstrate a typical query that is appropriate to use the geographic distance model. The example query we are using is SFO (San Francisco) to PIT (Pittsburgh) in the period of 07/19/2014 – 08/12/2014. The clustering result is showing below in Fig.7 with a Silhouette Score of 0.51.

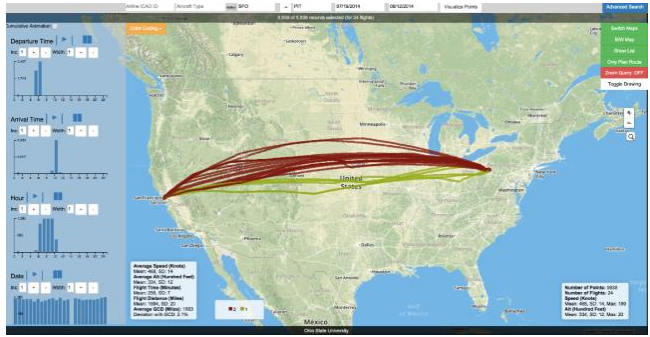

*Figure 7: Clustering result of SFO-PIT, Geographical Distance, Silhouette Score: 0.51*

However, if we used the cosine similarity model to cluster the same query, the result would not be as good as the geographic distance one. The cosine similarity model fails to find appropriate clustering, due to the highly similar directional vectors of each of the flights visualized, as illustrated in Fig. 8 with a Silhouette Score of only 0.39.

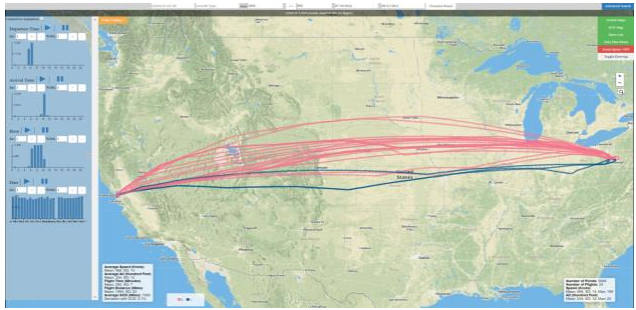

*Figure 8: Clustering result of SFO-PIT, Cosine Similarity, Silhouette Score: 0.39*

### 2) Example where Cosine Similarity Model is Preferred

In some cases, clustering using geographical distance may also fail to achieve desired results. An example of this is illustrated in Fig. 9 depicting flights between CMH and Philadelphia (PHL) from 10/1/2014 to 10/8/2014. Fig. 9 (left) illustrates a single cluster of flights when looked at the entirety of a flight, while (right) it is clearly seen that the arrival patterns into PHL follow two distinct paths (an easterly approach and a westerly approach).

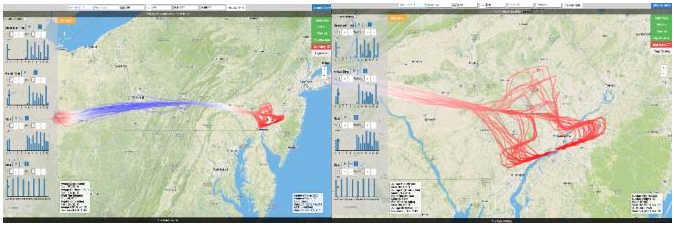

*Figure 9: CMH-PHL 10/1/2014 - 10/8/2014, entire flight path (left), arrival paths into PHL (right)*

In the case of CMH-PHL, the spatial-based methodology fails to perform this clustering as illustrated in Fig. 10. This is due to the fact that throughout the flight, the relative distance between the routes is so small that the algorithm fails to pick up the relatively small variations in the approach paths relative to the entire flight.

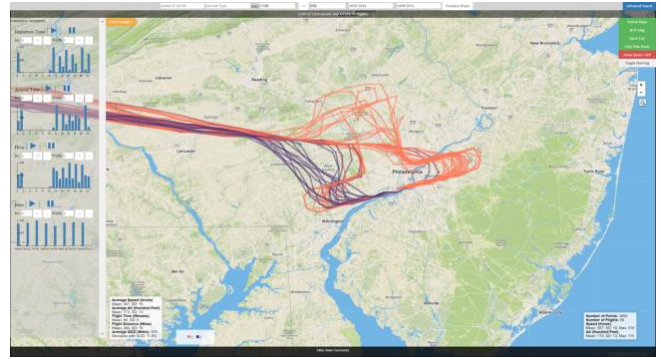

*Figure 10: Clustering result of CMH-PHL, Geographical distance, Silhouette Score: 0.46*

The Cosine Similarity model, however, generates a perfect clustering of these two approaches. As illustrated in Fig. 11, by using the Silhouette Score to find the statistically optimal threshold to the cosine similarity model, results in these categories of approaches.

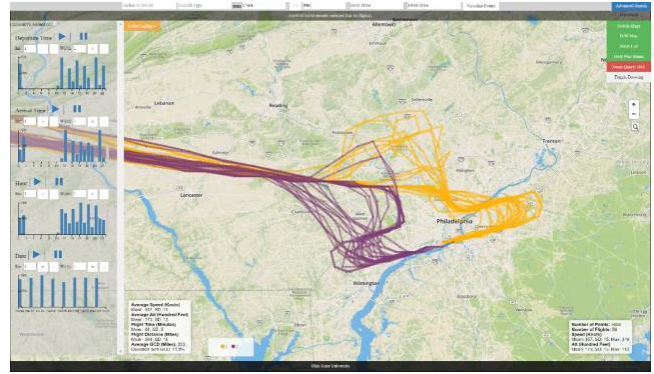

*Figure 11: Clustering Result of CMH-PHL arrivals into PHL, Cosine Similarity, Silhouette Score: 0.61*

### 3) Discussion of Model Selection

From the above examples, we may conclude that the Geographical Distance model is more accurate when clustering path groups that are significantly far away from each other though they may follow very similar shape or direction; the Cosine Similarity model, however, is more sensitive to directional differences but cannot distinguish if the flights are in the same direction. More generally, air flights often "near" each other during the departure and arrival phase of flight, but the difference between flight directions may be large to enter their planned flight routes. Therefore, the cosine similarity model may yield a good clustering result during the departure/arrival period just as the CMH-PHL example. For overall flights, and particularly clustering for enroute phases of flight, geographical distance is the more appropriate choice.

## V. METHODS TO IMPROVE MODEL PERFORMANCE

### A. Adding flexibility with Human-in-the-loop Clustering

Clustering using the Silhouette method always yields the optimal clusters statistically and hence will always provide the same result for a given set of flights. However, different users may have different intents in using the clustering function: some would like to detect anomalies by treating most flights in one big "normal" cluster while some would like to do a detailed analysis on flight segments expecting a more diverse clustering result. In addition, there might exist other possible clustering results that not "optimal" but still acceptable, and have only slightly lower Silhouette scores. For these reasons, our application also provides another option for users to manually enter a threshold to define the cutting level of the dendrogram. The users can then compare different results based on different thresholds which increases the flexibility of our application and provide opportunities for users to investigate different clustering experiments. This may be considered a "human-in-the-loop" clustering process.

In this next example, we demonstrate different clustering results and their corresponding statistical information by changing thresholds. We will use the Geographical Distance model in our example and change different distance thresholds to compare the clustering results and statistics information.

The example query we are using is the SFO-PIT one from the previous section. As illustrated in Fig. 7, the Silhouette method provides the result with 2 clusters and this result can be generated by applying a threshold of 80nm. However, by decreasing the threshold to 50nm, the geographical distance model eventually generates three significant path groups from north to south, which seems to have a better visualization match as illustrated in Fig.12. Tables 2 and 3 demonstrate the statistics for the two-cluster and three-cluster results respectively.

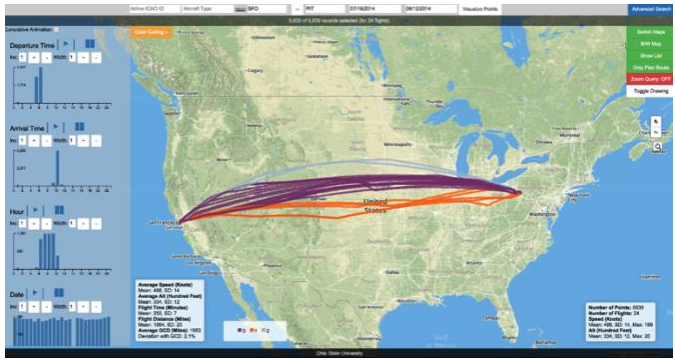

*Figure 12: Clustering result of SFO-PIT Geographical Distance threshold 50 nm - 3 cluster result*

*Table 2: Descriptive Statistics - SFO-PIT Geographical Distance, cluster threshold 80 nm*

| Cluster 1 | | Cluster 2 | |
|---|---|---|---|
| Number of Points: 4437 | Number of Flights: 19 | Number of Points: 1502 | Number of Flights: 5 |
| **Speed (Knots)** | | **Speed (Knots)** | |
| Mean: 465 | SD: 11 | Mean: 478 | SD: 18 |
| **Altitude (Hundred Feet)** | | **Altitude (Hundred Feet)** | |
| Mean: 337 | SD: 12 | Mean: 322 | SD: 6 |
| **Flight Distance (nm)** | | **Flight Distance (nm)** | |
| Mean: 1990 | SD: 25 | Mean: 1997 | SD: 17 |
| **Deviation with GCD** | | **Deviation with GCD** | |
| 1.9% | | 2.3% | |

*Table 3: Descriptive Statistics - SFO-PIT Geographical Distance, cluster threshold 50 nm*

| Cluster 1 | | Cluster 2 | | Cluster 3 | |
|---|---|---|---|---|---|
| Number of Points: 3971 | Number of Flights: 17 | Number of Points: 1502 | Number of Flights: 5 | Number of Points: 466 | Number of Flights: 2 |
| **Speed (Knots)** | | **Speed (Knots)** | | **Speed (Knots)** | |
| Mean: 465 | SD: 12 | Mean: 478 | SD: 18 | Mean: 463 | SD: 2 |
| **Altitude (Hundred Feet)** | | **Altitude (Hundred Feet)** | | **Altitude (Hundred Feet)** | |
| Mean: 338 | SD: 12 | Mean: 322 | SD: 6 | Mean: 332 | SD: 2 |
| **Flight Distance (nm)** | | **Flight Distance (nm)** | | **Flight Distance (nm)** | |
| Mean: 1990 | SD: 24 | Mean: 1999 | SD: 20 | Mean: 1994 | SD: 8 |
| **Deviation with GCD** | | **Deviation with GCD** | | **Deviation with GCD** | |
| 1.9% | | 2.4% | | 2.1% | |

The above example demonstrates the flexibility of the human-in-the-loop process. By changing different thresholds, the users are able to generate the desired clustering results. By decreasing the threshold, the agglomerative hierarchical clustering algorithm can further divide the existing clusters. For example, when the threshold is decreased from 80nm (suggested by the Silhouette method) to 50 nm, the above example detects one group of two flights from the original Cluster 1 as having a greater deviation from GCD and categorizes these two flights to a separate cluster.

The human-in-the-loop process may also be applied to the cosine similarity model with the user determining the cosine similarity threshold.

### B. Improving computing performance with Point Extraction

In our application, each record in the data is one flight path that contains multiple points consisting of latitude and longitude pairs. Based on the large number of data points in each flight, to save computing time, our application extracts a small portion of data points from each flight path – picking one point from every N points. Previous works, such as those previously referenced in [4] and [7] have also found that point extraction decreases significantly computation requirements in the clustering function while preserving the original relevant information, such as shape and direction, and eventually reduces noise.

Here, we have experimented with different N values – 1(keep every point, no extraction), 2, 4, 8, and 16. We will show some actual examples of how different N values would affect the clustering result and computation performance.

The first example still comes from the SFO (San Francisco) – PIT (Pittsburgh) flights. This query includes 265 flights and 39,839 flight points in total, which is a relatively large query.

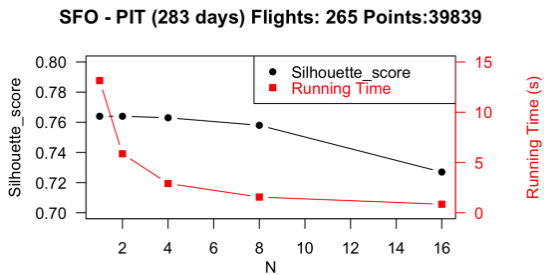

SFO - PIT (283 days) Flights: 265 Points:39839

*Figure 13: Experiment result of point extraction for SFO-PIT example*

Fig. 13 demonstrates that the clustering performance, as indicated by Silhouette Score, remains almost unchanged with points extraction until the value of N is 8. Moreover, the running time decreases significantly. Therefore, point extraction can help improve computing speed and maintain the same level of clustering performance.

However, this scenario may be different for other queries. The second example also comes from the previous query: CMH (Columbus) – PHL (Philadelphia). This query includes 58 flights and 3,650 flight points in total, which is a relatively small query.

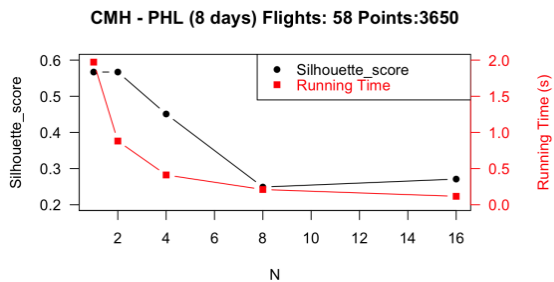

CMH - PHL (8 days) Flights: 58 Points:3650

*Figure 14: Experiment result of point extraction for CMH-PHL example*

Fig. 14 demonstrates that the clustering performance remains the same when N is increased from 1 to 2, but the Silhouette score starts to decrease significantly if we continue to increase the value of N.

The above are two typical examples and we have made multiple experiments on point extraction. We found that for the query whose flights are generally long-distance (more points per flight) and have fewer directional changes, the clustering function could still maintain the same level of performance even using a small proportion of data points. However, for the query whose flights are generally short-distance (fewer points per flight) and have frequent directional changes, the performance of clustering is more sensitive to the point extraction technique. But a small value of N can still improve the computing performance while providing a reasonable Silhouette score. Therefore, we would suggest using a large value of N (e.g. 8 or 16) for a long-distance flight query and using a small value of N (e.g. 2 or 4) for a short-distance flight query.

## VI. CONCLUSIONS AND FUTURE WORK

Using FAA TFMS data and the *DV8* data visualization tool, two clustering models were developed to categorize groups of flights based on their flight paths by applying a hierarchical clustering algorithm and Silhouette Score metric to find optimal clustering results. Specifically, a spatial-based model based on geographic distance between routes, and a vector-based model based on cosine similarity were developed and tested over three examples. The results found that while in some cases both models produced reasonable results, the spatial-based model performed better for entire flight routes, particularly all traveling over long distances in the same general direction, while the vector-based model performed better in particular phases of flight where physical distance between aircraft may be close, but directionality is more varied, as in the case of various approaches to an airport.

To take advantage of both cosine similarity's good performance during the departure/arrival period and geographic distance's good performance during the enroute flying period, future work includes a plan to segment flight routes into three segments, departure, enroute, and arrival, and apply the appropriate clustering algorithm to each segment, creating a more diverse collection of clusters per route. Other advanced but computationally expensive distance functions to compute trajectory similarity would also be implemented, such as Fréchet distance [16], for experiments and comparison. *DV8* is a framework tool related to secondary development. Finally, we will also consider flights in a 3-dimensional space by including flight altitude in the route clustering algorithms.

In addition, we have proposed a "human-in-the-loop" option as an alternative of the automatic clustering to increase flexibility for users that may have different clustering intents. We have also demonstrated that the point extraction technique can significantly improve the computing efficiency while maintaining the same clustering performance. With the point extraction technique, the fast-time visualization capabilities of *DV8* enhances the capability of interactively analyzing large numbers of flights within few seconds or even sub-seconds. We will continue to develop more ML capabilities within the *DV8* framework, using current TFMS, as well as other available air traffic data feeds. These capabilities will further enhance the tools required to fully evaluate the performance of flight routes in the national airspace system.